\documentclass[reprint, prx,amsmath,amssymb,showpacs,floatfix,longbibliography]{revtex4-1}
\usepackage[breaklinks=true,colorlinks,citecolor=blue,linkcolor=blue,urlcolor=blue]{hyperref}
\usepackage{graphicx,epsfig}
\usepackage{amsmath}
\usepackage{xcolor}
\usepackage{color}
\usepackage{bm}
\usepackage{cases}
\usepackage{dsfont}
\renewcommand{\BibitemShut}[1]{}

\def\be{\begin{equation}}
\def\ee{\end{equation}}
\def\nn{\nonumber}
\def\bea{\begin{eqnarray}}
\def\eea{\end{eqnarray}}

\begin{document}

\title{Chiral anomalies induced transport in Weyl semimetals in quantizing magnetic field}
\author{Kamal Das}
\email{kamaldas@iitk.ac.in}
\author{Sahil Kumar Singh}
\email{sahilks137@gmail.com}
\author{Amit Agarwal}
\email{amitag@iitk.ac.in}
\affiliation{Department of Physics, Indian Institute of Technology Kanpur, Kanpur 208016, India}

\begin{abstract}
Weyl semimetals host relativistic chiral quasiparticles, which display quantum anomalies in the presence of external electromagnetic fields. Here, we study the manifestations of chiral anomalies in the longitudinal and planar magneto-transport coefficients of Weyl semimetals, in the presence of a quantizing magnetic field. We present a general framework for calculating all the transport coefficients in the regime where multiple Landau levels are occupied. We explicitly show that all the longitudinal and planar transport coefficients show Shubnikov-de Haas like quantum oscillations which are periodic in $1/B$. Our calculations recover the quadratic-$B$ dependence in the semiclassical regime, and predict a linear-$B$ dependence in the ultra-quantum limit for all the transport coefficients. 
\end{abstract}

\maketitle

\section{Introduction} 

Weyl semimetals (WSMs) host massless chiral relativistic quasiparticles which show very interesting and novel phenomena \cite{Wan11,Burkov11,Armitage18,Kim13,Yang14,Burkov14,Parameswaran14,Burkov15,Cortijo15,Chernodub18,Song19,Xiang19,Sonowal19,Wang20,Sadhukhan20}. One of the most interesting aspects of a massless relativistic chiral fluids, in quantum field theory, is the breakdown of the chiral gauge symmetry in presence of an external electromagnetic field~\cite{Adler69, Bell69,Nielsen83}. This results in the chiral anomaly (CA) which manifests in the non-conservation of chiral charge~\cite{Adler69, Bell69,Nielsen83,Landsteiner16}. A condensed matter realization of this was first explored in the lattice theory of Weyl fermions by Nielson and Ninomiya in 1983~[\onlinecite{Nielsen83}]. They predicted that the CA will give rise to a positive longitudinal magneto-conductivity, which is linear in the magnetic field strength ($B$), for ultra-high magnetic field in the diffusive limit. With the recent realizations of WSM, there have been several experiments which report positive magneto-conductivity or negative magneto-resistance and attribute it to the CA \cite{Wan11,Burkov11,Armitage18,Kim13,Xiong15,Huang15a,Li16a, Zhang16}. 

Relativistic chiral fluids in a gravitational field also display the mixed chiral-gravitational anomaly, which results in non-conservation of the chiral energy~\cite{Landsteiner11,Lucas16}. This manifests in the magneto-thermal experiments in the form of positive magneto-thermopower and positive magneto-thermal conductivity~\cite{Lucas16, Gooth17, Stone18, Das19c}, both of which have also been observed in recent experiments~[\onlinecite{Hirschberger16, Jia16}]. In addition to their manifestations in longitudinal magneto-transport, CAs have also been shown to give rise to the planar Hall effects in all transport coefficients [\onlinecite{Burkov17, Nandy17, Das19b, Sharma19, Nandy19, Li_Hui18, Kumar18, Li_P18, Yang19,Das19c}]. Recently, we predicted another anomaly, the thermal chiral anomaly in which a temperature gradient collinear with the magnetic field gives rise to charge and energy imbalance between the opposite chirality Weyl fermions \cite{Das19c}. 

\begin{figure}[t]
\includegraphics[width = \linewidth]{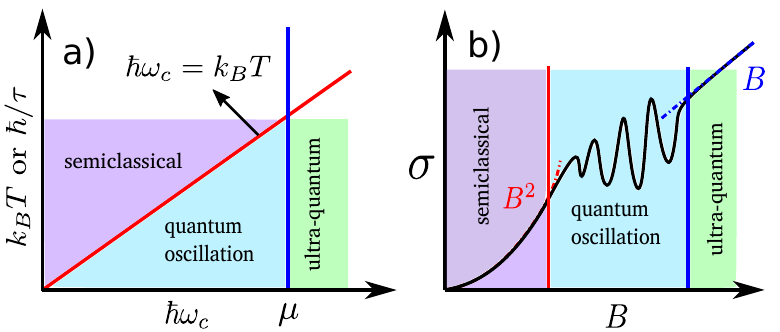}
\caption{(a) The three different magneto-transport regimes. 
In the semiclassical regime (purple area) multiple LL are occupied, but the impurity scattering or thermal smearing makes them indistinguishable ($\hbar \omega_c < k_B T ~{\rm or}~\hbar/\tau)$. In the quantum oscillation regime (blue), multiple LL are occupied and they are distinguishable, while in the ultra-quantum regime (green), only the lowest LL is occupied ($\hbar \omega_c > \mu$).
(b) The magnetic field dependence of the magneto-conductance in the three transport regimes for a WSM. 
\label{fig_1}}
\end{figure}

However, the bulk of the theoretical work till date has been focussed on the semiclassical transport regime which displays a quadratic-$B$ dependence of all transport coefficients [\onlinecite{Son13,Kim14a, Burkov14, Yip15, Burkov15, Das19a, Lundgren14, Kim14b, Spivak16, Sharma16, Das19b}], or only in the ultra-quantum regime~[\onlinecite{Nielsen83}] (see Fig.~\ref{fig_1}). Recently, there has been a prediction of quantum oscillations in the longitudinal magneto-conductivity~[\onlinecite{Gorbar14, Deng19a}]. So a natural question to ask is, what happens to the other magneto-transport coefficients? Furthermore, can all the three distinct transport regimes, highlighted in Fig.~\ref{fig_1}, be explored within a unified framework?

In this paper, we attempt to answer these and other related questions. We present a general framework for calculating all the transport coefficients in the regime where multiple Landau levels are occupied and connect them to the different CAs. We explicitly show that all the longitudinal and planar transport coefficients show Shubnikov-de Haas (SdH) like quantum oscillations which are periodic in $1/B$. Our calculations recover the quadratic-$B$ dependence in the semiclassical regime, and predict a linear-$B$ dependence in the ultra-quantum limit for all the magneto-transport coefficients.
The rest of the paper is organized as follows: In Sec.~\ref{quantization} we discuss the Landau quatization in WSMs, and in Sec.~\ref{anomalies} we connect the anomalous magnet-transport in WSMs to different CAs. In Sec.~\ref{BTF} we present the general formalism of calculating all the CA induced magneto-transport coefficients. In Sec.~\ref{lngtdnl} we calculate the longitudinal magneto-transport coefficients, followed by the exploration of the  magneto-transport coefficients in the planar Hall setup in Sec.~\ref{planar}. We discuss the experimental possibilities in Sec.~\ref{discussion}, and summarize our results in Sec.~\ref{cnclsn}. 

\section{Landau quantization in WSM} 
\label{quantization} 

\begin{figure}[t]
\includegraphics[width = \linewidth]{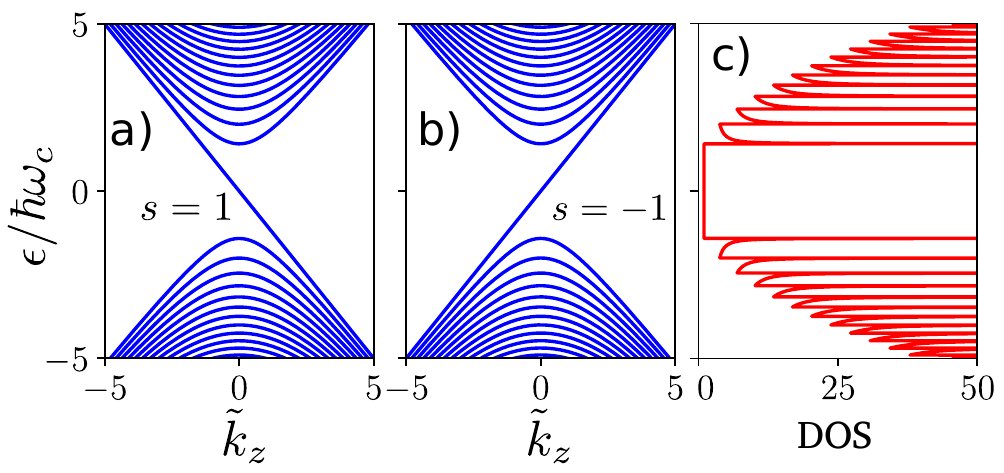}
\caption{The Landau level energy spectrum in  a WSM for (a) positive and (b) negative chirality Weyl node, respectively. The lowest LLs disperse linearly and are chiral in nature. (c) The corresponding density of states (DOS)
of the LL spectrum. The DOS is scaled by ${\frak D}/(h v_F)$. The DOS for the lowest ($n=0$) LL is constant, and it increases as the energy moves away from the Weyl nodes. The oscillation in the DOS will also manifest in all the transport coefficients.}
\label{fig_2}
\end{figure}

In presence of magnetic field, the Hamiltonian of a single Weyl cone (of chirality $s$), after Peierls substitution, is given by 
\be \label{ham_1}
\hat H^s = s v_F {\bm \sigma}\cdot(\hat{\bf p} + e {\bf A})~.
\ee
Here, $-e$ is the electronic charge, $v_F$ is the Fermi velocity, ${\bm \sigma} = (\sigma_x, \sigma_y, \sigma_z)$ is a vector composed of the three Pauli matrices, and $\bf A$ is the vector potential corresponding to the magnetic field ${\bf B} = \nabla \times {\bf A}$.
Considering the magnetic field along the $z$-direction and using the Landau gauge with ${\bf A} = (-By, 0, 0)$, it is straight forward to calculate the energy spectrum of the Hamiltonian in Eq.~\eqref{ham_1}. The Landau level (LL) energy spectrum is given by, 
\be \label{LL_1}
\epsilon_n^s =
\begin{cases}
-s \hbar v_F k_z & n = 0 \\
 \pm  \sqrt{(\hbar v_F k_z)^2 + 2 n (\hbar \omega_c)^2} & n \geq 1~.
\end{cases}
\ee
Here, $n$ denotes the LL index and we have defined the cyclotron frequency, $\omega_c = v_F/l_B$ with $l_B = \sqrt{\hbar/(e|B|)}$ being the magnetic length scale. 
In the rest of the manuscript we will use $B = |{\bf B}|$. 
The energy spectrum of Eq.~\eqref{LL_1} is shown in Fig.~\ref{fig_2}(a) and (b). We emphasize that the lowest LL are chiral in nature, {\it i.e.}, right (left) movers for negative (positive) chirality node, and will play a crucial role in the CAs~[\onlinecite{Nielsen83}] discussed in this paper. In contrast, all the $n \geq 1$ LLs are achiral (support both right and left movers), and they play an important role in quantum (SdH like) oscillations. 

Each of the LLs is highly degenerate and the degeneracy is specified by $\frak{D} = 1/2\pi l_B^2$. The DOS of the LL spectrum is shown in Fig.~\ref{fig_2}(c). 
The group velocity of the quasi-particles in these LLs is given by
\be 
v_{nz}^s = \dfrac{\partial \epsilon_n^s}{\hbar \partial k_z} =
\begin{cases}
-s v_F & n = 0 \\
\hbar v_F^2 k_z/ \epsilon_n^s & n \geq 1~.
\end{cases}
\ee
The carriers of the lowest LLs have a constant velocity, and are either left movers (for $s = 1$) or right movers (for $s = -1$). 
Next, we discuss the origin of quantum anomalies from the lowest chiral LLs, and explore their manifestation in electric and thermal magneto-transport experiments in WSM~[\onlinecite{Landsteiner11, Son13, Gooth17, Schindler18, Das19c}].

\section{Equilibrium currents and coefficients of chiral anomalies}
\label{anomalies}

One simple way to understand the origin of CAs in WSM is to calculate the equilibrium (no externally applied bias voltage or temperature gradient) current, in presence of a magnetic field. 
The equilibrium charge and energy current for each Weyl node is given by $\{j_{e, {\rm eq}}^s, j_{{\cal E}, {\rm eq}}^s \}= {\frak D}\sum_n \int \frac{dk_z}{2 \pi} v_{nz}^s \{- e, \epsilon_n^s\}f_n^s $. Here, $f_n^s$ is the Fermi-Dirac distribution function for the $n$-th LL: $f_n^s \equiv 1/[1 + e^{\beta\left(\epsilon_n^s - \mu\right)}]$ with $\beta \equiv 1/k_B T$, and $\mu$ denotes the chemical potential. 
We evaluate these using the Sommerfeld expansion [\onlinecite{Stone18, Das19c}] to obtain,
\bea \label{eq_chrg_cur}
j_{e, {\rm eq}}^s &=& -e \left(\mu {\cal C}_0^s  + k_B T {\cal C}_1^s \right)B~,
\\ \label{eq_ener_cur}
j_{\cal E, {\rm eq}}^s &=& \left(\mu^2\frac{{\cal C}_0^s}{2}+\mu k_BT {\cal C}_1^s + k_B^2T^2 \dfrac{{\cal C}_2^s}{2} \right)B~.
\eea
Here, the coefficients ${\cal C}_\nu^s$ for $\nu = \{0,1,2\}$, are given by 
\be \label{c_anm}
{\cal C}_\nu^s =\dfrac{e}{2 \pi \hbar} \sum_n \int \dfrac{d k_z}{2\pi} v_{nz}^s \left(\dfrac{\epsilon_n^s - \mu}{k_B T}\right)^\nu \left(-\dfrac{\partial f_n^s}{\partial \epsilon_n^s} \right)~.
\ee
In Eq.~\eqref{c_anm}, only the chiral lowest LLs ($n=0$) contribute to the sum and we have ${\cal C}_\nu^s \propto s$,  the chirality of the Weyl node. Thus, the existence of the chiral LLs is an essential ingredient to obtain the non-zero charge and energy currents for each node even in equilibrium. These chiral currents are in turn related to the CAs in WSMs~[\onlinecite{Gooth17, Schindler18,Das19c}]. We had earlier derived equations similar to Eqs.~\eqref{eq_chrg_cur} and \eqref{eq_ener_cur} in the semiclassical regime, where the Berry curvature of the Weyl nodes  played an important role~[\onlinecite{Das19c}].

\begin{figure}[t]
\includegraphics[width = \linewidth]{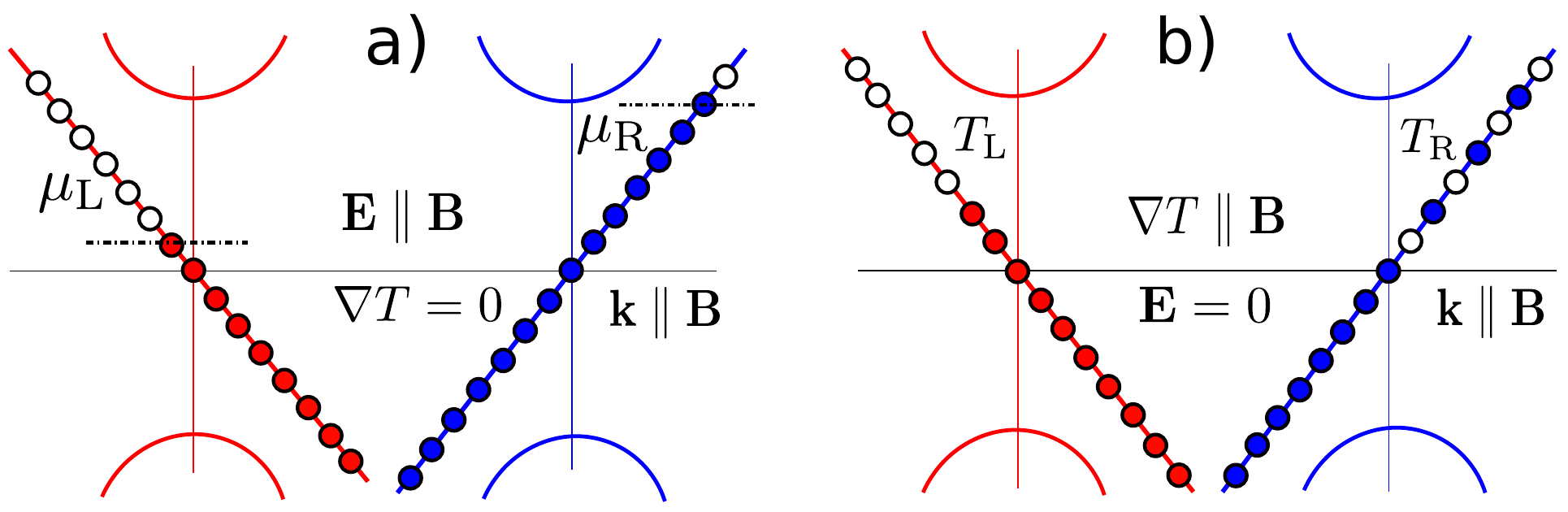}
\caption{Chiral anomalies induced charge and energy imbalance in a WSM in the ultra-quantum limit. (a) The electrical chiral anomaly induced non-conservation of chiral charge in presence of parallel electric and magnetic fields.  (b) The  mixed chiral-gravitational anomaly induced non-conservation of chiral energy in presence of a temperature gradient parallel to the magnetic field.}
\label{fig_4}
\end{figure}

The quantities ${\cal C}_0^s$, ${\cal C}_2^s$ and ${\cal C}_1^s$ are known as the coefficient of the chiral (or axial) anomaly~[\onlinecite{Son13}], the coefficient of the mixed chiral- (or axial-) gravitational anomaly~[\onlinecite{Landsteiner11, Gooth17, Schindler18}] and the coefficient of the thermal chiral (or axial) anomaly~[\onlinecite{Das19c}], respectively. 
We emphasize that the coefficient ${\cal C}_1^s$ vanishes in the $\beta \mu \to \infty$ limit, and has been relatively unexplored.
Evaluating Eq.~\eqref{c_anm}, 
we obtain, 
\begin{subnumcases}{{\cal C}_{\nu}^s = -s \dfrac{e}{4 \pi^2 \hbar^2} \label{c_anm_1}}
{\cal F}_0(\beta \mu)  & $\nu = 0$ \\
{\cal F}_1(\beta \mu)   & $\nu = 1$ \\
{\cal F}_2 (\beta \mu) & $\nu = 2$~.
\end{subnumcases}
Here, we have defined ${\cal F}_0(x)=\frac{1}{1+e^{-x}}$ for the chiral anomaly, 
${\cal F}_1(x)=\frac{x}{1+e^{x}}+\ln(1+e^{-x})$ for the thermal chiral anomaly and 
${\cal F}_2(x)=\frac{\pi^2}{3}+2\text{Li}_2(-e^{-x})-2x\ln(1+e^{-x})-\frac{x^2}{1+e^x}$ for the mixed chiral-gravitational anomaly. The $x$-dependence of these functions is explicitly shown in Fig.~\ref{fig_4}. In the $x \rightarrow \infty$ limit, these functions are constant [See. Eq.~\eqref{hi}], with ${\cal F}_2$ being the largest of the three and ${\cal F}_1 \to 0$. However, the thermal chiral anomaly coefficient ${\cal F}_1$ has a finite value for finite $x$. 

In equilibrium, the total charge and energy current from all the Weyl nodes in a WSM adds upto zero, as opposite chirality nodes always appear in pairs in a WSM. However, the chiral charge current ($ j_{e, {\rm eq}}^+ - j_{e, {\rm eq}}^-$) and energy current ($j_{\cal E, {\rm eq}}^+ - j_{\cal E, {\rm eq}}^-$) are non-zero even in equilibrium.  More interestingly, in presence of an eternal electric field or a temperature gradient, this leads to charge and energy imbalance between pair of opposite chirality Weyl nodes. 
Below, we explore the consequence of this in magneto-transport experiments. 


\begin{figure}[t]
\includegraphics[width =0.84\linewidth]{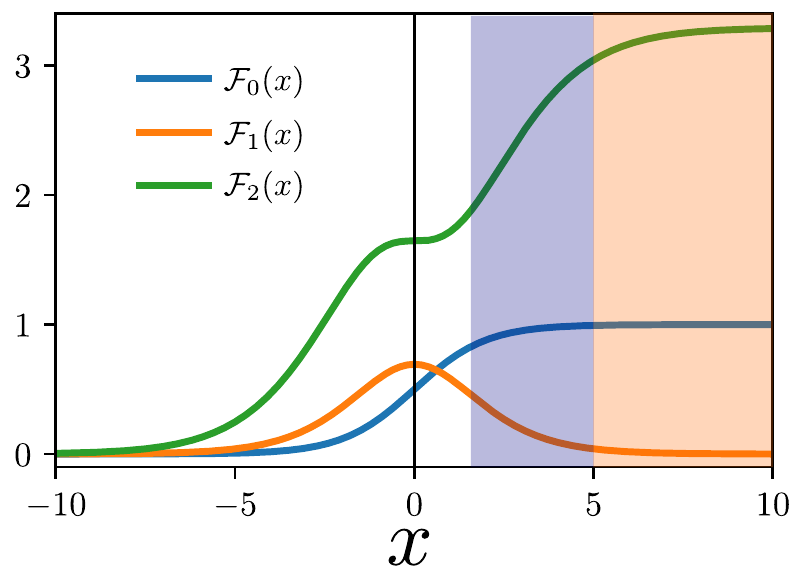}
\caption{The temperature dependence ($x = \beta \mu$) of the dimensionless functions associated with the CAs.  
Note that in $x \rightarrow \infty$ or $T \to 0$ for a finite $\mu >0$ limit, these functions are constant as shown in the orange shaded region. In this region, the coefficient of thermal chiral anomaly ${\cal F}_1 \to 0$. However, it has a finite value 
in the finite $x$ regime. For this paper, we will mostly focus on the regime $\mu > k_B T$ or $x>1$ which is shaded in blue and, in the $\mu \gg k_B T$ regime shaded in orange. 
\label{fig_3}}
\end{figure}

\section{Non-equilibrium currents and chiral anomalies} 
\label{BTF}

An applied electric field or temperature gradient drives the system out of equilibrium. In the Boltzmann transport formalism the non-equilibrium distribution function (NDF), $g_n^s$, within the relaxation time approximation, satisfies the following equation [\onlinecite{Deng19a, Das19c}], 
%
\be \label{bte_1}
\partial_t g_n^s + \dot {\bf k}_n^s \cdot {\bm \nabla}_{\bf k} g_n^s + \dot {\bf r}_n^s \cdot {\bm \nabla}_{\bf r} g_n^s = -\dfrac{g_n^s - \bar g_n^s }{\tau} - \dfrac{\bar g_n^s - f_{n}^s}{\tau_v}~.
\ee
Here, $\bar g_n^s$ is the {\it local equilibrium} distribution function considered to be the Fermi function for the $n$th LL with a node dependent chemical potential $\mu^s$ and temperature $T^s$. 
The first term in the right hand side of Eq.~\eqref{bte_1} represents the relaxation of the NDF to the local equilibrium through the intra-node scattering rate $1/\tau$. 
The intra-node scattering does not alter the number of carriers in the respective node, and its impact is similar to that in other metals as well. In contrast, the second term represents 
the inter-node scattering with a relaxation rate of $1/\tau_v$, which attempts to undo the impact of the chiral imbalance  [Eq.~\eqref{cont_1_f}-\eqref{cont_2_f}] and restore a steady state carrier distribution function. 
In a typical WSM with broken time reversal symmetry, the Weyl nodes are separated in the momentum space. If we are in a regime of small $\mu$ so that the Fermi wave-vector is smaller than the 
separation of the Weyl nodes, then we have $\tau_v \gg \tau$~[\onlinecite{Burkov15}]. 

The idea of a steady state in presence of CAs and inter-node scattering becomes more evident from the continuity equations of particle number and heat density for electric field and temperature gradient applied along the direction of magnetic field. 
Integrating Eq.~\eqref{bte_1} over all the states in a single cone, we obtain the particle number conservation equation (within the linear response) to be 
\be \label{cont_1_f}
\dfrac{\partial {\cal N}^s}{\partial t} +{\bm \nabla}_{\bf r} \cdot {\bf J}^s_n + eE B {\cal C}_0^s   =  - \dfrac{{\cal N}^s - {\cal N}_0^s}{\tau_v}~.
\ee
Here, ${\bm \nabla}_{\bf r} \cdot {\bf J}^s_n = k_B {\cal C}_1^s B \nabla T$ is the divergence of the particle current. 
The quantities $\{{\cal N}_0^s,{\cal N}^s\} = \frak{D} \sum_n \int \frac{dk_z}{2 \pi}\{f_{n}^s,g_n^s\}$ 
are the total particle number density in each Weyl cone before and after applying external fields, respectively. In Eq.~\eqref{cont_1_f}, the terms with ${\cal C}_0^s EB$ and ${\cal C}_1^s B \nabla T$ represent  CAs induced flow of particle~[\onlinecite{Das19c}].

Following a similar procedure, we construct the continuity equation for heat density. Multiplying Eq.~\eqref{bte_1} by $\tilde \epsilon_n^s \equiv \epsilon_n^s-\mu$ and integrating over all the states of a given Weyl node, we obtain 
\be \label{cont_2_f}
\dfrac{\partial {\cal Q}^s}{\partial t} + {\bm \nabla}_{\bf r} \cdot {\bf J}^s_Q + eE B {\cal C}_1^s k_B T   =   - \dfrac{{\cal Q}^s - {\cal Q}_0^s}{\tau_v}~.
\ee
Here, ${\bm \nabla}_{\bf r} \cdot {\bf J}^s_Q = k_B^2 {\cal C}_2^s T B \nabla T$ is the divergence of the heat current. The quantities $\{{\cal Q}_0^s,{\cal Q}^s\} = \frak{D} \sum_n \int \frac{dk_z}{2 \pi}\tilde \epsilon_n^s \{f_{n}^s,g_n^s\}$ 
are the heat density in each cone, before and after applying external fields, respectively. In Eq.~\eqref{cont_2_f}, $EB {\cal C}_1^s$ and ${\cal C}_2^s B \nabla T$ represent the CAs induced flow  of the heat density[\onlinecite{Das19c}].
In constructing Eqs.~\eqref{cont_1_f}-\eqref{cont_2_f}, we have used the fact that the inter-node scattering does not change the number of carrier and energy of 
each Weyl node. 

To evaluate the charge and heat currents in this steady state, we now calculate the NDF from Eq.~\eqref{bte_1}. Within the linear response regime, it is reasonable to expect that the change in chemical potential and temperature in each cone is small: 
$\delta \mu^s \equiv \mu^s -\mu < \mu$ and $\delta T^s \equiv T^s -T < T$. Then to the lowest order in $\delta \mu^s$ and $\delta T^s$, the NDF can be expressed as 
\bea \label{dstr_f_1}\nn
g_n^s &=& f_{n}^s - \tau v_{nz}^s \left(e  E + \dfrac{\epsilon_n^s - \mu}{T} \nabla T\right)\left(-\dfrac{\partial f_n^s}{\partial \epsilon_n^s} \right)  
\\
&+& \left(1 - \dfrac{\tau}{\tau_v}\right)\left( \delta \mu^s +\dfrac{\epsilon_n^s - \mu}{T} \delta T^s  \right)\left(-\dfrac{\partial f_n^s}{\partial \epsilon_n^s} \right).
\eea
In Eq.~\eqref{dstr_f_1}, the chiral chemical potential $\delta \mu^s$ and chiral temperature $\delta T^s$ are given by 
\be \label{chrl_mu_T}
\begin{pmatrix}
\delta \mu^s \\
k_B \delta T^s
\end{pmatrix}
=-\tau_v B~
[{\cal D}^s]^{-1}
\begin{pmatrix}
{\cal C}_0^s & {\cal C}_1^s\\
{\cal C}_1^s & {\cal C}_2^s
\end{pmatrix}
\begin{pmatrix}
eE \\
k_B \nabla T 
\end{pmatrix}~.
\ee
Here, we have defined  the magnetic field dependent generalized energy density matrix, ${\cal D}^s \equiv \begin{pmatrix}
{\cal D}_0^s & {\cal D}_1^s\\
{\cal D}_1^s & {\cal D}_2^s
\end{pmatrix},$ with
\be \label{DOS}
{\cal D}_\nu^s =  {\frak D}\sum_n \int \dfrac{d k_z}{2\pi} \left( \dfrac{\epsilon_n^s - \mu}{k_B T}\right)^\nu \left(-\dfrac{\partial f_n^s}{\partial \epsilon_n^s} \right),
\ee 
and $\nu = \{0,1,2\}$.  


%
Equation~\eqref{chrl_mu_T} quantifies the chiral charge and energy imbalance in the two Weyl nodes. These imbalances are proportional to the coefficients of CAs and inversely proportional to the generalized energy density. 
The former is due to the fact that CAs are responsible for the charge and heat imbalances. The latter is a consequence of the fact that a smaller DOS will lead to a larger change in $\delta \mu^s$ and $\delta T^s$ and vice versa. 
These imbalances are schematically depicted in Fig.~\ref{fig_3} in the ultra-quantum limit with only the lowest LLs being occupied. Figure~\ref{fig_3}(a) shows the $\delta \mu^s$ induced by the electrical chiral anomaly and Fig.~\ref{fig_3}(b) displays the $\delta T^s$ generated by the mixed chiral-gravitational anomaly. In addition to these two anomalies, the thermal chiral anomaly has a non-zero contribution to both the charge and energy imbalances, though this contribution is relatively smaller~\cite{Das19c}.

\section{Longitudinal magneto-transport}
Having obtained the NDF in the Landau quantization regime, 
we now proceed to calculate the magneto-transport coefficients. The steady state non-equilibrium charge and heat current for each Weyl node is defined as 
$\{j_e^s, j_Q^s\} = \sum_n \frak{D} \int \frac{dk_z}{2 \pi} v_{nz}^s  \{-e,  \tilde \epsilon_n^s  \}g_n^s$. 
%
Focussing only on the anomaly induced contribution which is proportional to the inter-node scattering time $\tau_v$, we obtain 
\small
\be \label{currents_1}
\begin{pmatrix} 
j_e^s \\
j_Q^s
\end{pmatrix}
= - \tau_v  B^2
\begin{pmatrix}
-e {\cal C}_0^s & -e {\cal C}_1^s \\
k_B T{\cal C}_1^s &  k_B T{\cal C}_2^s
\end{pmatrix}
[{\cal D}^s]^{-1}
\begin{pmatrix}
{\cal C}_0^s & {\cal C}_1^s \\
{\cal C}_1^s & {\cal C}_2^s
\end{pmatrix}
\begin{pmatrix}
eE \\
k_B \nabla T 
\end{pmatrix}.
\ee
\normalsize
Note that in Eq.~\eqref{currents_1}, the magnetic field dependence of the charge and heat current comes from i) the $B^2$ term, ii) the DOS which depends on $B$ via the LL spectrum, and iii) magnetic field dependence of $\tau_v$. 

The transport coefficients can now be easily obtained from the phenomenological relations for linear response: $j_{e,i} = \sum_j [\sigma_{ij}E_j - \alpha_{ij}\nabla_j T$] and $ j_{Q,i} = \sum_j [{\bar \alpha}_{ij}E_j - {\bar \kappa}_{ij}\nabla_j T ]$. 
Here, $\sigma$, $\alpha$, $\bar\alpha$ and $\bar \kappa$ denote the electrical, thermo-electric, electro-thermal and constant voltage thermal conductivity matrix, respectively. In the limiting case of $\beta \mu \to \infty$ (or $\mu \gg k_B T$), the thermal chiral anomaly coefficient ${\cal C}_1^s \to 0$. 
In the same limit, we find ${\cal D}_0^s, {\cal D}_2^s \gg {\cal D}_1^s$. Using these, Eq.~\eqref{currents_1} can be rewritten as  
%
%
\small
\be \label{simple_cur}
\begin{pmatrix} 
j_e^s \\
j_Q^s
\end{pmatrix}
= \tau_v  B^2
\begin{pmatrix}
\frac{1}{{\cal D}_0^s}({e\cal C}_0^s)^2  & \frac{{\cal D}_1^s}{{\cal D}_0^s {\cal D}_2^s}e{\cal C}_0^sk_B{\cal C}_2^s \\
 T \frac{{\cal D}_1^s}{{\cal D}_0^s {\cal D}_2^s}e{\cal C}_0^sk_B{\cal C}_2^s & T \frac{1}{{\cal D}_2^s}(k_B{\cal C}_2^s)^2
\end{pmatrix}
\begin{pmatrix}
E \\
- \nabla T 
\end{pmatrix}.
\ee 
\normalsize
From Eq.~\eqref{simple_cur}, we note that $\sigma \propto ({\cal C}_0^s)^2$, ${\bar \kappa \propto ({\cal C}_2^s)^2}$, $\alpha \propto {\bar \alpha} \propto{\cal C}_0^s {\cal C}_2^s$. Thus, it is reasonable to associate different transport coefficients with different CAs.
The presence of thermal chiral anomaly coefficient in the more general Eq.~\eqref{currents_1} leads to a more complicated dependence of the transport coefficients on the anomaly coefficients.

We now explore the different regimes of magneto-transport depending on the strength of applied magnetic field: 
i) the ultra-quantum regime where only the lowest ($n=0$) LLs are occupied, ii) the quantum oscillation regime with a few distinguishable LLs being occupied, and iii) the semiclassical regime with many but undistinguishable LLs being occupied. 

\label{lngtdnl}

\subsection{Ultra-quantum regime}
In the ultra-quantum regime, only the lowest ($n=0$) LL is occupied and we have $\hbar \omega_c \gg\mu$. 
In this regime, the CA coefficients have already been calculated in Eq.~\eqref{c_anm_1}. We calculate the finite temperature DOS and its energy moments defined in Eq.~\eqref{DOS}, to be 
\begin{subnumcases}{{\cal D}_{\nu,0}^s = \dfrac{\frak D}{2 \pi \hbar v_F} \label{DOS_0} }
{\cal F}_0(\beta \mu) & $\nu = 0$\\
{\cal F}_1(\beta \mu) & $\nu = 1$\\
{\cal F}_2 (\beta \mu) & $\nu = 2$~.
\end{subnumcases}
Note that similar to the coefficient of thermal CA, ${\cal C}_{1}^s$, the first energy moment of the DOS, ${\cal D}_{1,0}^s$ is relatively smaller than the other two.
Using these expressions in Eq.~\eqref{currents_1}, we obtain the anomaly induced charge current to be
\be \label{chrg_cur_LLL_1}
j_e^s =\dfrac{e}{h} \frac{e B \tau_v v_F}{h}  \left[e{\cal F}_0(\beta\mu) E    + k_B{\cal F}_1(\beta\mu) \nabla T  \right]~.
\ee
We highlight the linear-$B$ dependence of the charge current. 
The corresponding charge conductivity is given by $\sigma^s_{zz} = \frac{e^2}{ 4 \pi^2 \hbar} \frac{ v_F  \tau_v}{l_B^2}{\cal F}_0(\beta \mu)$.
Since ${\cal F}_0(x \to \infty) \approx 1$, this reproduces the previously obtained results~[\onlinecite{Nielsen83, Aji12, Son13, Gorbar14, Zhang16a}] in the limit $\beta \mu \to \infty$. 

The corresponding thermoelectric conductivity is given by $\alpha_{zz}^s  = - \frac{k_B e}{ 4 \pi^2 \hbar} \frac{ v_F  \tau_v}{l_B^2}{\cal F}_1(\beta \mu)$
which is small but non-zero for finite $\beta \mu$. Remarkably, we find that the magnetic field induced thermoelectric conductivity is negative, unlike its semiclassical counterpart~\cite{Spivak16, Das19c} and violates the Mott relation.
However, in the limiting case of $\beta \mu \to \infty$, $\alpha_{zz}^s \to 0$, consistent with the results of Ref.~[\onlinecite{Gooth17,Schindler18}]. As a consequence of this vanishing $\alpha$, the magnetic field induced change in the Seebeck coefficient, $S = \alpha \sigma^{-1}$, is equal to the magnetic field induced change in $\sigma$. This in turn implies that the magnetoresistance (MR) in the Seebeck coefficient is equal to the MR in the resistivity. This is in contrast with the semiclassical regime where the MR in the Seebeck coefficient is twice that of the MR in the resistivity~\cite{Spivak16, Das19c, Jia16}.

The CA induced heat current is  obtained to be
\be \label{heat_cur_LLL_1}
j^s_{Q}=\dfrac{k_B T}{h}\frac{eB v_F\tau_v }{h}  \left[-e {\cal F}_1(\beta\mu)E  - k_B  {\cal F}_2(\beta\mu)\nabla T\right]~.
\ee
Similar to the charge current, the heat current also shows a linear-$B$ dependence. The validity of the Onsager's reciprocity relation, $\bar \alpha({\bf B}) = T \alpha(-{\bf B})$ can also be confirmed. The constant voltage thermal conductivity is given by 
\be \label{kappa}
\bar \kappa_{zz}^s  = \frac{k_B^2 T}{ 4 \pi^2 \hbar} \frac{ v_F  \tau_v}{l_B^2}{\cal F}_2(\beta \mu)~.
\ee 
Interestingly, for finite $\beta \mu$, we find that $\bar \kappa_{zz}^s$ and $\sigma_{zz}^s$ are not connected via the Wiedemann-Franz law. However, the Wiedemann-Franz law gets restored in the $\beta \mu \to \infty$ limit, and we have $\bar \kappa_{zz}^s = \frac{\pi^2}{3} (\frac{k_B}{e})^2\sigma^s_{zz}$.
This linear-$B$ dependence of the thermal conductivity in the ultra-quantum limit has also been observed in recent experiments~[\onlinecite{Schindler18, Vu19}]. 
\begin{figure}[t]
\includegraphics[width = \linewidth]{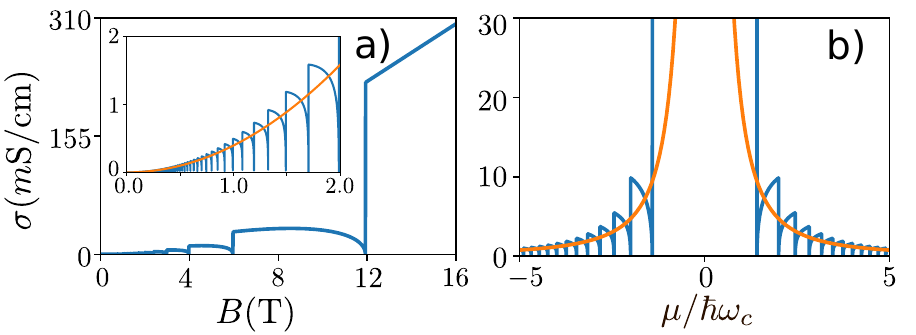}
\caption{(a) The variation of the longitudinal charge conductivity with $B$ (for $\mu = 25$ meV). Note the linear-$B$ dependence in the ultra-quantum regime. The inset shows the low field behaviour and the semiclassical quadratic-$B$ dependence is shown in orange color. (b) The $\mu$ dependence of the longitudinal charge conductivity (for $B = 2$~T). Note that the spurious divergence of the longitudinal conductivity as $\mu \to 0$ in the semiclassical limit, is not there in the LL picture. Here, we have chosen $v_F = 2 \times 10^5$~m/s and $\tau_v = 10^{-9}$~s. 
\label{fig_5}}
\end{figure}
\subsection{Quantum oscillation regime}
\label{mul_LL}
In this section, we will discuss the scenario when multiple LLs are occupied. For this, the chemical potential has to be larger than the separation between the LLs ($\mu > \hbar \omega_c$). Since we want to highlight the oscillations in the magneto-transport coefficients, we work in the regime $ k_B T \ll \hbar \omega_c $, so that the temperature broadening does not smear out the signatures of the discrete LLs. These two conditions combine to yield, $\mu \gg k_B T$ and thus we focus on the limit of 
$\beta\mu \to \infty$ in this section. 
In this limit, the coefficients of CAs are given by
\begin{subnumcases}{{\cal C}_{\nu}^s = -s \dfrac{e}{4 \pi^2 \hbar^2} \label{hi}}
1       &   $\nu = 0$ \\
0       &   $\nu = 1$ \\
\pi^2/3 &   $\nu = 2$~.
\end{subnumcases} 
%
%

In contrast to the anomaly coefficients, the DOS and its energy moments get contributions from all the filled LLs. 
The highest occupied LL index can be calculated to be $n_c = {\rm int} [\mu^2/(2 \hbar^2 \omega_c^2)]$, and we obtain 
\begin{subnumcases}{{\cal D}_{\nu}^s = 
\dfrac{\frak{D}}{2 \pi \hbar v_F}\label{DOS_1} }
\Theta_0 & $\nu =0$ \label{DOS_1_1}\\
\dfrac{2\pi^2}{3\beta \mu}\Theta_1 & $\nu =1$ \\
\dfrac{\pi^2}{3}\Theta_0 & $\nu =2$~.
\end{subnumcases}
Here, we have defined $\Theta_0 \equiv 1+\sum_{n = 1}^{n_c}2/\lambda_n $ and $\Theta_1 \equiv -\sum_{n = 1}^{n_c}2n (\hbar \omega_c/\mu)^2/\lambda_n^3$ with $\lambda_n =\sqrt{1 - 2 n(\hbar \omega_c/\mu)^2}$. Note that $n_c \propto {\rm int}[1/B]$.
So the number of occupied LLs is inversely proportional to $B$. This combined with Eq.~\eqref{DOS_1_1} is what leads to 
SdH like oscillations in the longitudinal magneto-conductivity with a period proportional to $1/B$. As a consistency check, we note that Eq.~\eqref{DOS_1_1} obtained here, is identical to Eq.~(4) of Ref.~[\onlinecite{Deng19a}].
%
%
We find that while ${\cal D}_0^s$ and ${\cal D}_2^s$ are more or less temperature independent (for $\mu> k_B T$), 
${\cal D}_1^s$ depends inversely on $\beta \mu$ and vanishes in the limit $T \to 0$. 
\begin{figure}[t]
\includegraphics[width = \linewidth]{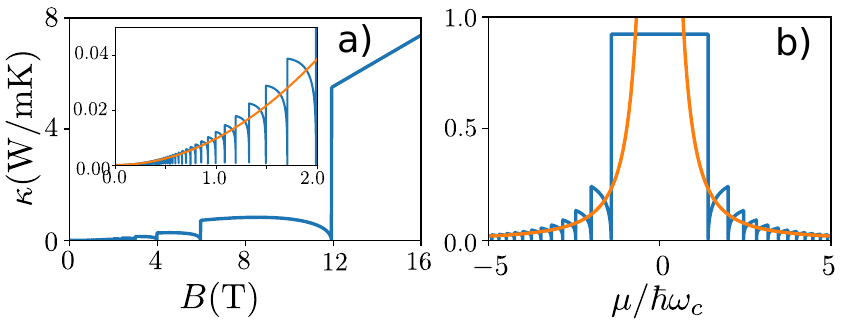}
\caption{(a) The variation of the longitudinal thermal conductivity with $B$. The linear-$B$ dependence in the ultra-quantum regime (for $B>12$ T here) is evident. The inset shows the small $B$ behaviour and the semiclassical $B^2$ dependence is shown by the orange line. (b) The $\mu$ dependence of the longitudinal thermal conductivity. The parameters are identical to those of Fig.~\ref{fig_5} and $T = 10$~K.}
\label{fig_6}
\end{figure}
Using Eq.~\eqref{hi} and \eqref{DOS_1} in Eq.~\eqref{simple_cur}, we calculate the charge current to be, 
\be \label{chrg_cur_mul_LL_1}
j_e^s = \dfrac{e}{h} \dfrac{eB \tau_v v_F}{h} \left( \dfrac{e}{\Theta_0} E - k_B \dfrac{2\pi^2}{3\beta \mu} \dfrac{\Theta_1}{\Theta_0^2}  \nabla T\right).
\ee 
The $1/\Theta_0$ term in the charge conductivity originates from the $1/{\cal D}_0^s$ term in Eq.~\eqref{simple_cur}, and it gives rise to SdH like oscillations in the longitudinal conductivity which are periodic in $1/B$ \cite{Gorbar14, Deng19a}. The $B$ dependence of the longitudinal conductivity is displayed in Fig.~\ref{fig_5}. 
%
Additionally, we find that the longitudinal thermoelectric conductivity also shows SdH like quantum oscillations arising from the discreteness of the LLs. In this case the $\Theta_1/\Theta_0^2$ term in Eq.~\eqref{chrg_cur_mul_LL_1} governs the features of the quantum oscillations. The Onsager's reciprocity relation, $\bar \alpha ({\bf B}) = T \alpha (-{\bf B})$ holds for the longitudinal thermoelectric conductivity, even in the quantum oscillation regime. 

We calculate the heat current to be
\be \label{heat_cur_mul_LL_1}
j_Q^s = \dfrac{k_B T}{h} \dfrac{eB \tau_v v_F}{h} \dfrac{\pi^2}{3}\left( \dfrac{2e}{\beta \mu} \dfrac{\Theta_1}{\Theta_0^2}  E - \dfrac{k_B}{\Theta_0}  \nabla T\right).
\ee
Equations~\eqref{chrg_cur_mul_LL_1} and Eq.~\eqref{heat_cur_mul_LL_1} are the main results of this paper. 
We have demonstrated that the longitudinal thermal conductivity also possess SdH like quantum oscillations with features very similar to that of the charge conductivity (both dictated by $1/\Theta_0$.)
The dependence of the thermal conductivity on the magnetic field and the chemical potential, is shown in Fig.~\ref{fig_6}. 

\begin{figure}[t]
\includegraphics[width = \linewidth]{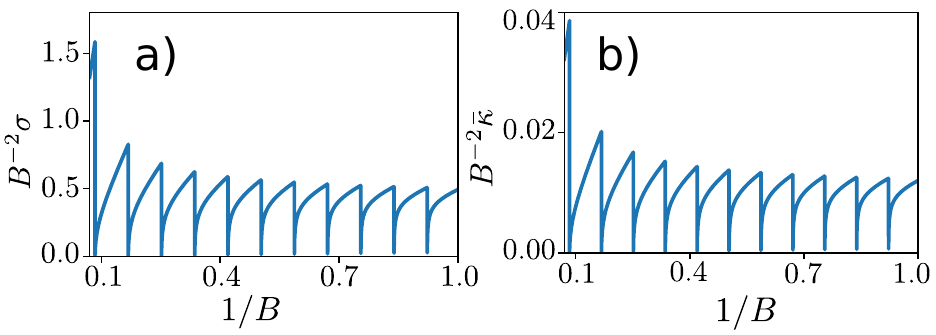}
\caption{The (a) charge and (b) thermal conductivity as a function of the inverse of the magnetic field. The constant period of oscillation in $1/B$ is evident. The period of oscillation is determined by Onsager's quantization rule defined in Eq.~\eqref{period_B}.}
\label{fig_7}
\end{figure}

Remarkably, we find that the period of oscillations of the longitudinal transport coefficients $\sigma$ and ${\bar \kappa}$, defined in Eqs.~\eqref{chrg_cur_mul_LL_1}-\eqref{heat_cur_mul_LL_1} satisfies the Onsager's quantization rule for SdH oscillations~\cite{Ashcroft76} in the transverse conductivity. However, this is not surprising, considering the fact that for both of these the origin lies in the DOS of the LLs. Onsagar's quantization rule for the SdH oscillations in the transverse conductivity states that the period of conductivity oscillation (in $1/B$) is given by, 
\be \label{period_B}
\Delta\left(\dfrac{1}{B}\right) = \dfrac{2 \pi e}{\hbar} \dfrac{1}{A_e}~.
\ee
Here, $A_e$ is the extremal cross section of the Fermi surface in a plane perpendicular to the magnetic field. 
For our longitudinal conductivity, we find that both the charge and thermal conductivity vanishes at $\mu^2 = 2n (\hbar \omega_c)^2$. This yields the period of oscillation to be $\Delta_1(1/B) = 2 e \hbar v_F^2/\mu^2$. And since the Fermi surface 
in an isotropic WSM, is spherical with $A_e = \pi k_F^2$, $\Delta_1(1/B)$ is consistent with Eq.~\eqref{period_B}. The SdH like oscillations in $1/B$ in the longitudinal components of $\sigma$ and $\bar \kappa$ are explicitly shown in Fig.~\ref{fig_7}. 
\subsection{Semiclassical regime}
In this section, we show that for small $B$ when many LLs are occupied, we can recover the semiclassical results for the 
CA induced transport coefficients \cite{Das19c}. 
For closely spaced LLs such that $\mu \gg \hbar \omega_c$, we can replace the discrete sum over LLs by an integral: 
$\sum_0^{n_c} \to \int_{0}^{n_c} dn$.
In this limit, we obtain $\Theta_0 \approx 2 (\mu/\hbar \omega_c)^2$ and $\Theta_1 \approx 2 (\mu/\hbar \omega_c)^2$. Using these expressions it is straight forward to calculate 
\begin{subnumcases}{{\cal D}_{\nu}^s \approx 
\dfrac{\mu^2}{2 \pi^2}\dfrac{1}{\hbar^3 v_F^3}\label{DOS_2} }
1 & $\nu =0$ \label{DOS_1_2}\\
\dfrac{2\pi^2}{3\beta \mu} & $\nu =1$ \\
\dfrac{\pi^2}{3} & $\nu =2$~.
\end{subnumcases}
These expressions are identical to the DOS derived in Ref.~[\onlinecite{Das19c}].

Now, the charge current can be obtained to be 
%
\be \label{chrg_cur_SCT}
j_e^s = \dfrac{e^2}{8 \pi^2 \hbar} \dfrac{(eB)^2 v_F^2}{\mu^2}\tau_v v_F \left( E - \dfrac{k_B}{e} \dfrac{2\pi^2}{3\mu \beta}   \nabla T\right).
\ee
The charge conductivity is identical to the previously reported results~[\onlinecite{Son13, Spivak16, Das19c}] which shows quadratic-$B$ and positive magneto-conductivity. This semiclassical quadratic-$B$ dependence is a well-established signature of the CA in the low field limit, and it has also been experimentally verified~[\onlinecite{Xiong15,Huang15a,Li16a}]. The thermoelectric conductivity is also consistent with the previously obtained semiclassical results~[\onlinecite{Spivak16, Das19c}], and with the experimental observations in Dirac semimetals~[\onlinecite{Jia16,Hirschberger16}].
A similar calculation yields the heat current to be, 
\be \label{heat_cur_SCT}
j_Q^s = \dfrac{k_B T}{8 \pi^2 \hbar} \dfrac{(eB)^2 v_F^2}{\mu^2} \tau_v v_F\dfrac{\pi^2}{3}\left(\dfrac{2e}{\beta \mu}  E - k_B \nabla T\right).
\ee
We note that the thermal conductivity obtained here, is identical to the previous reports~[\onlinecite{Spivak16, Das19c}] and it has recently been measured in GdPtBi~[\onlinecite{Schindler18}]. The magnetic field and the Fermi energy dependence of charge conductivity in this regime, is shown in In Fig.~\ref{fig_5} on top of the quantized LL results. In Fig.~\ref{fig_5}(a), the three different transport regimes, with different $B$-dependence, are evident. In the inset, the yellow line shows the semiclassical fitting. In Fig.~\ref{fig_5}(b) the Fermi energy dependence is displayed. In Fig.~\ref{fig_6} we have shown the same for the thermal conductivity.

\section{Planar Hall effects}
\label{planar}
So far we have explored the impact of CAs in the longitudinal magneto-transport coefficients. 
However, it has been shown that the origin of planar Hall effects and anisotropic longitudinal transport coefficients in non-magnetic materials can also be related to CAs~[\onlinecite{Burkov17, Nandy17, Das19b, Sharma19, Nandy19}].
Here, we explore the impact of quantized LLs, on all the planar Hall transport coefficients, and explore the possibility of SdH like quantum oscillations in them. 

\begin{figure}[t]
\includegraphics[width = \linewidth]{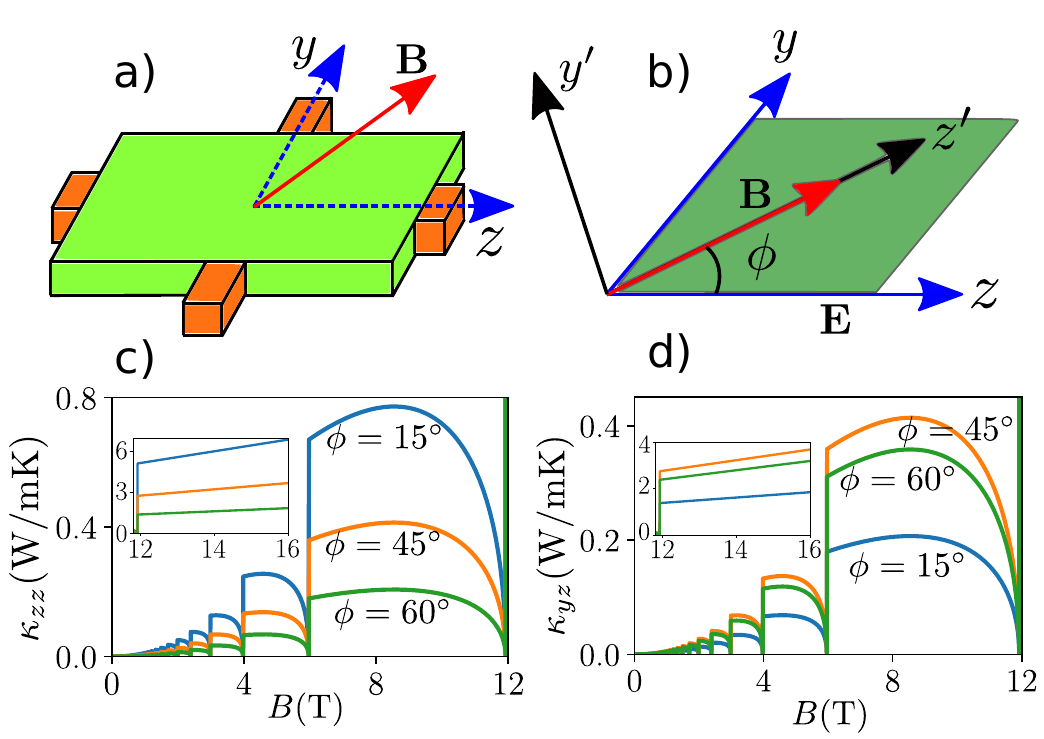}
\caption{(a) A schematic of the planar Hall geometry experiments. (b) For ease of calculations, we work in the frame of reference with an axis aligned along the magnetic field. This is rotated with respect to the laboratory frame (aligned along the electric field or temperature gradient), by an angle $\phi$. 
%
The $B$-dependence of a) the anisotropic longitudinal thermal conductivity ($\bar \kappa_{zz} \propto \cos^2\phi$) and b) the planar Righi-Leduc effect ($\bar \kappa_{yz} \propto \cos\phi \sin\phi$) for different angles between the magnetic field and the temperature gradient. The inset shows the  contribution of the lowest LL. 
\label{fig_8}}
\end{figure}

In the planar Hall setup~\cite{Li_Hui18, Kumar18, Li_P18, Yang19}, we measure the longitudinal and Hall transport coefficients in the plane of the ${\bf E}-{\bf B}$ (or ${\bm \nabla} T-{\bf B}$) fields, as shown in Fig.~\ref{fig_8}(a). Here the electric field is applied along the $z$-direction with a planar magnetic field applied at an angle $\phi$, so that ${\bf B} = B\left(\cos \phi \hat{\bf z} + \sin \phi \hat{\bf y} \right)$. 
%
However, it turns out that the calculations for the case where the electric field or temperature gradient is parallel or 
perpendicular to the magnetic field are relatively easier~[\onlinecite{Lukose07, Alisultanov17}]. Thus, we perform our calculations in a rotated frame of reference ($z^\prime$-$ y^\prime$), so that the magnetic field lies along the $z^\prime$-axis of the new frame, as shown in Fig.~\ref{fig_8}(b). The coordinates in the two frames are related as, 
\be 
\begin{pmatrix}
z^\prime \\
y^\prime
\end{pmatrix}
=
\begin{pmatrix}
\cos \phi & \sin \phi \\
-\sin \phi & \cos \phi
\end{pmatrix}
\begin{pmatrix}
z \\
y
\end{pmatrix}.
\ee
If the transport coefficients are denoted by $L_{ij}^\prime$ in this rotated frame, then the transport coefficients in the lab frame, $L_{ij} = \{\sigma_{ij},\alpha_{ij}, \bar{\alpha}_{ij}, \kappa_{ij}\}$, are given by 
%
\small
\be  \label{trnsf_TC}
\begin{pmatrix}
L_{zz} & L_{zy}\\
L_{yz} & L_{yy}
\end{pmatrix}
=
\begin{pmatrix}
\cos \phi & -\sin \phi \\
\sin \phi & \cos \phi
\end{pmatrix}
\begin{pmatrix}
L_{zz}^\prime & L_{zy}^\prime\\
L_{yz}^\prime & L_{yy}^\prime
\end{pmatrix}
\begin{pmatrix}
\cos \phi & \sin \phi \\
-\sin \phi & \cos \phi
\end{pmatrix}.
\ee
\normalsize
%


For an electric field applied at an angle to the magnetic field, there are two different effects at play. The parallel component of the electric field (the component parallel to the magnetic field: $E_\parallel = E\cos \phi$) makes the crystal momentum time dependent: $k_z^\prime \to k_z + e E_\parallel t/\hbar$. This modifies the NDF of the electronic states as shown in Eq.~\eqref{dstr_f_1}, with the substitution $E \to E_{\parallel}$. 
The perpendicular component of the electric field ($E_\perp = E \sin\phi$) modifies the LL spectrum and the associated DOS, as derived in detail in Appendix \ref{crsd_fld}. 
%
The modified LL spectrum gives rise to a nonlinear (in $E_\perp$) DOS as shown in Eq.~\eqref{DOS_2}. A similar approach was used in Ref.~[\onlinecite{Deng19b}] to demonstrate that the longitudinal planar conductivity has a $\cos^6 \phi$ angular dependence, and the planar Hall component has a $\cos^5 \phi \sin \phi$ angular dependence, arising from the non-linear terms in the DOS. However, in this work we work in the linear response regime, and focus on the quantum oscillation of the planar thermal transport coefficients. 
Using the LL dispersion, we calculate the velocity along the magnetic field (to zeroth order in $E_{\perp}$) to be, 
\be
v_{n z^\prime} = {v_F}  \left(s_n \dfrac{k_z}{\sqrt{\frac{2 n}{\tilde l_B^2} + k_z^2}} - s \delta_{n,0} \right)~. 
\ee
Here, $s_n = sgn(n)$. 
There is also a Lorentz velocity component along the $x-$direction (perpendicular to the ${\bf B}-{\bf E}$ plane), $v_{nx} = {E}\sin\phi/B$, which gives rise to the conventional Hall effect.  The velocity component along the $y'$ direction is zero. 
%
 %

Now, it is straight forward to calculate the transport coefficients in the rotated frame of reference, $z^\prime$-$y^\prime$.  
Using Eq.~\eqref{trnsf_TC} to revert back to the laboratory frame, we find that the $B$ dependent part of the transport coefficients pick up the familiar angular dependence given by 
\bea \label{L_zz}
\{\sigma_{zz},\alpha_{zz}, \bar{\alpha}_{zz}, \bar \kappa_{zz}\}  &=& \{\sigma_{zz}^\prime,\alpha_{zz}^\prime, \bar{\alpha}_{zz}^\prime, \bar \kappa_{zz}^\prime\}  \cos^2 \phi, \\ \label{L_zy}
\{\sigma_{yz},\alpha_{yz}, \bar{\alpha}_{yz}, \bar \kappa_{yz}\}  &=& \{\sigma_{zz}^\prime,\alpha_{zz}^\prime, \bar{\alpha}_{zz}^\prime, \bar \kappa_{zz}^\prime\} \sin\phi \cos\phi. \nonumber \\
\eea
In deriving Eqs.~\eqref{L_zz} and \eqref{L_zy}, we have used the fact that $L_{yy}^\prime = L_{yz}^\prime =0$. This follows from the fact that the velocity component along the $y^\prime$ axis is zero. This clearly establishes that the planar Hall transport coefficients retain the $B$-dependence of the longitudinal transport coefficients in the linear response regime. Consequently, they also display the three regimes of i) ultra-quantum transport which is linear in $B$, ii) SdH like quantum oscillation transport regime, and iii) the semiclassical transport regime which is quadratic in $B$. 


The $B$ dependence of the longitudinal and the planar (Righi-Leduc) thermal conductivity is shown in Fig.~\ref{fig_8}(c) and (d), for different orientations of the magnetic field. The quantum oscillations in both of these can be clearly seen. Similar quantum oscillations will also be there in all other transport coefficients. 

\section{Discussion}
\label{discussion}

In this paper, we have presented all calculations within the constant scattering time approximation. The energy and temperature dependence of the scattering timescale can be easily included and it does not change the qualitative features of the discussed transport coefficients. However, we have to be more careful in analyzing the magnetic field dependence of the scattering timescale. For the short range impurity scattering, realized via neutral defects, the scattering rate is proportional to the density of states~[\onlinecite{Burkov15,Aji12,Lu17}]. 
Now, in the ultra-quantum limit the DOS is proportional to $B$, and hence $\tau_v \propto 1/B$. Thus, the transport coefficients may become completely independent of the magnetic field in this regime. For the case of multiple filled LLs, the magnetic field dependence of the scattering timescale is more complicated. Additionally, the magnetic field dependence for other scattering mechanisms like charged impurities and phonons amongst others is still an open problem. 

As shown in Fig.~\ref{fig_1}, the semiclassical or quantum transport regimes are quantified by $\omega_c \tau$ or $\beta \hbar\omega_c$. Now, if $\omega_c \tau \gg 1$ along with $\beta \hbar\omega_c \gg 1$, then the LL broadening caused either by the impurities or by the thermal smearing is less than the LL separation, making them distinguishable. This is the regime where the discreteness of 
the LLs manifests itself in all physical quantities including transport coefficients. Here, the shortest scattering time-scale dominates the disorder induced broadening. This typically turns out to be the intra-node scattering time ($\tau$) in WSMs. For a system with $v_F = 2 \times 10^5$~m/s and $\tau \approx 10^{-12}$~s we have $B \gg 0.02$~T for observing quantized Landau levels in transport experiments. So a value of $B \approx 1$~T is very likely to show the SdH oscillation regime. This corresponds to $\hbar \omega_c \approx 5~m$eV, which is easily much larger than the thermal energy scale for temperatures $T < 60$~K.
Furthermore, the use of Boltzmann transport equation assumes the presence of a weak disorder strength which preserves the shape of the Fermi surface. This is specified via the condition $\mu \gg \hbar/\tau$. Thus, to explore the ultra-quantum limit, we have to be in the regime $\hbar/\tau < \mu < \hbar \omega_c$. For a magnetic field of $B = 10$ T and $\tau \approx 10^{-12}$ s, this translates to $0.7~ {\rm meV} < \mu < 16$~meV.

\section{Conclusions}
\label{cnclsn}
The origin of longitudinal magneto-resistance is physically very intriguing since the electrons do not feel any Lorentz force along the applied electric field. Particularly, in chiral fluids such as WSM the longitudinal transport coefficients can also originate from CAs which makes longitudinal transport in WSM even more exciting \cite{Pal10, Gao17}. In this paper, we have presented a unified framework for calculating the transport coefficients which captures all three transport regimes: i) ultra-quantum, ii) quantum oscillation, and iii) semiclassical. We derive explicit analytical expressions for all the CAs induced magneto-transport coefficients in the regime of multiple LL being occupied. We explicitly show that the mixed chiral-gravitational anomaly induced longitudinal thermal conductivity, and the planar Righi-Ludec conductivity, display SdH like oscillations with features similar to that in the longitudinal conductivity. Additionally, we find a linear-$B$ dependence of {\it all} magneto-transport coefficients in the ultra-quantum limit, while a quadratic-$B$ dependence in the semiclassical regime, and SdH like oscillations in the intermediate regime. 
Our work will be useful in analyzing and interpreting the exciting magneto-transport experiments in WSMs. 

\appendix

\section{Crossed electric and magnetic fields} 
\label{crsd_fld}
The configuration of crossed fields where magnetic field is applied perpendicular to the bias voltage is common in the context of classical Hall effect. In this section, we solve the LL spectrum for such scenario. Lets consider the magnetic field to be $B \hat {\bf z}$ and electric field to be $E \hat {\bf y}$. The corresponding electromagnetic potentials we choose as ${\bf A} = ( -By , 0, 0)$ and $\phi = -Ey$. 
Now, the LL spectrum can be obtained by solving the wave equation $\hat H\Psi = \epsilon \Psi$ with 
\be 
\hat H = v_F \left[ \sigma_x \left(\hat p_x  - eBy \right) + \sigma_y \hat p_y  + \sigma_z \hat p_z \right] + e E y\sigma_0.
\ee
Here, $\sigma_0$ is a $2 \times 2$ identity matrix. Following Ref.~[\onlinecite{Lukose07}], we can recast this wave equation in the four momentum language with $p_0 = \epsilon/v_F$ and make the Lorentz transformation as~[\onlinecite{Alisultanov17, Deng19b}]
\be 
\begin{pmatrix}
\hat p_0 \\
\hat p_1
\end{pmatrix}
=
\begin{pmatrix}
\cosh \theta & \sinh \theta \\
\sinh \theta & \cosh \theta
\end{pmatrix}
\begin{pmatrix}
\hat {\tilde p}_0 \\
\hat {\tilde p}_1
\end{pmatrix}.
\ee
Here, $ \tanh \theta =\frac{E}{v_F B}$. This Lorentz transformation, takes us to a new frame which moves along the positive $x$-direction with velocity $v_F\tanh \theta$. In this new frame, there is no electric field. 
Now, using the identities $\sigma_x^2 = \sigma_0$ and $\left(\sigma_0 \cosh \theta - \sigma_x \sinh \theta\right) = \exp(-\theta \sigma_x)$, we rewrite the wave-equation as
\be \label{wve_eqn_2}
v_F \left[ \sigma_x \left(\hat {\tilde p}_x -  \dfrac{ eB}{\cosh \theta}\tilde y\right)  + \sigma_y \hat {\tilde p}_2 + \sigma_z \hat {\tilde p}_z - \sigma_0 \hat {\tilde p}_0   \right] \tilde \Psi = 0. 
\ee
Here, $\tilde \Psi(\tilde t,\tilde x, \tilde y, \tilde z )= \exp(-\sigma_x \theta/2)\Psi(t, x,  y,  z )$.  Equation~\eqref{wve_eqn_2} represents the familiar wave equation in presence of a magnetic field only and its eigen-value and wave functions are known. The only difference is the strength of the magnetic field is modified as $B \to B/\cosh \theta$. Using $\tilde B = B/\cosh \theta$ and $s_n = sgn(n)$ the LL spectrum is obtained to be
\be
\tilde \epsilon_{n}(\tilde p_z)= s_n \sqrt{( v_F \tilde p_z)^{2} + 2 |n| \hbar e  \tilde B v_F^2} - s v_F \tilde p_z \delta_{n,0}~.
\ee
After doing the inverse Lorentz transformation, we obtain
\be  \label{energy_2}
\epsilon^s_n(k_x, k_z) = \dfrac{1}{\eta^2} \hbar v_F {\cal K}_n^s(k_z) + \hbar v_F k_x \tanh \theta~.
\ee
Here, we have defined $\eta^{-2} = \sqrt{1 - \tanh^2 \theta}$ and
\be
{\cal K}^s_n(k_z) = sgn(n) \sqrt{\frac{2 |n|}{(l_B \eta)^2}+ k_z^2} - s k_z \delta_{n,0}~.
\ee
The most notable effect of the perpendicular electric field is the second term in Eq.~\eqref{energy_2}. As a result of this, the carriers have finite velocity along the $x$-axis, perpendicular to the ${\bf E}$-${\bf B}$ plane. It is straight forward to calculate
\be
v_{n z} = \dfrac{v_F}{\eta^2}  \left(s_n \dfrac{k_z}{\sqrt{\frac{2 n}{\tilde l_B^2} + k_z^2}} - s \delta_{n,0} \right);~~v_{nx} = \dfrac{E}{B}~.
\ee
We note that the $x$-component of the velocity is simply the Lorentz velocity, which is a constant and identical for all the LLs. This is what gives rise to the classical Hall effect.

We emphasize that Eq.~\eqref{wve_eqn_2} represents a harmonic oscillator with center at $y = (\tilde l_B)^2 \tilde p_x/\hbar$. In the lab frame this can be written as
\be
y_c^\prime = l_B^2 kx +  l_B^2  {\cal K}_n^s(k_z) \sinh \theta~.
\ee
The perpendicular electric field in this expression lifts the degeneracy of LLs, consequently it modifies the DOS. Using the general formula of DOS, $\rho(\epsilon) =\sum_{n =0}^{n_c} \frac{1}{(2 \pi)^2} \iint dk_x  dk_z \delta (\epsilon_n^s - \epsilon)$, a small calculation yields 
\be \label{DOS_2}
\rho(\epsilon) = \eta^2 \rho_0 \left[2\sum_{n=0}^{n_c}  \dfrac{\lambda_n(\epsilon,+) -\lambda_n(\epsilon, -)}{eE L_y} - 1\right]~.
\ee
Here, we have defined 
\be \label{lambda}
\lambda_n(\epsilon, \pm) = \sqrt{\left(\epsilon \pm \dfrac{eEL_y}{2} \right)^2- 2|n| (\hbar \omega_c \eta^{-3})^2}~.
\ee
Equation~\eqref{DOS_2} is a non-linear function of the electric field strength.

In the limit, of $eEL_y \ll \epsilon$, expanding Eq.~\eqref{lambda} in powers of the strength of the electric field we obtain the DOS to be $\rho(\epsilon) = \eta^2 \rho_0 \Theta_0$. Here, 
\be 
\Theta_0 = 2\sum_{n=0}^{n_c}  \dfrac{1}{\sqrt{1 - 2|n| \left(\hbar \omega_c \eta^{-3}\right)^2}} - 1~.
\ee
This expression of the DOS was used in Ref.~[\onlinecite{Deng19b}] to calculate the angular dependence of the planar Hall conductivity. However, in this paper we will focus on the linear response regime, and retain only the term independent of the electric field. To the lowest order in $E$, we have $\eta \to 1$, and Eq.~\eqref{DOS_2} reduces to the expression of the DOS derived in Eq.~\eqref{DOS_1}, which was independent of $E$.

\bibliography{CA_LL_TT_v7.bib}

\end{document}